\documentclass[a4paper,fleqn,usenatbib,useAMS]{mnras}

\usepackage{graphicx}	% Including figure files
\usepackage{amsmath}	% Advanced maths commands
\usepackage{amssymb}	% Extra maths symbols
\usepackage{multicol}        % Multi-column entries in tables
\usepackage{bm}		% Bold maths symbols, including upright Greek
\usepackage{pdflscape}	% Landscape pages
\usepackage[encapsulated]{CJK}
\usepackage{ucs}
\usepackage[utf8x]{inputenc}
\usepackage{url}
\usepackage{hyperref}
\usepackage{breakurl}
\usepackage{color}

\usepackage[dvipsnames]{xcolor}

\newcommand{\kms}{\,km\,s$^{-1}$} % kilometres per second
\usepackage[T1]{fontenc}
\usepackage{ae,aecompl}

\usepackage{newtxtext,newtxmath}
\title[NGC\,7635 - testing the WBB theory]{The Bubble Nebula NGC\,7635 -- testing the wind-blown bubble theory}
\author[J.A.\,Toal\'{a} et al.]{J.A.\,Toal\'{a}$^{1}$\thanks{E-mail:j.toala@irya.unam.mx},
  M.A.\,Guerrero$^{2}$, H.\,Todt$^{3}$, L.\,Sabin$^{4}$, L.M.\,Oskinova$^{3}$, Y.-H.\,Chu$^{5}$, 
  \newauthor{G.\,Ramos-Larios$^{6}$ and V.M.A.\,G\'{o}mez-Gonz\'{a}lez$^{1}$}\\
$^{1}$Instituto de Radioastronom\'{i}a y Astrof\'{i}sica, IRyA-UNAM, Apartado postal 3-72, 58090, Morelia, Mich., Mexico\\
$^{2}$Instituto de Astrof\'{i}sica de Andaluc\'{i}a, IAA-CSIC, Glorieta de la Astronom\'{i}a s/n, 18008 Granada, Spain\\
$^{3}$Institut f\"{u}r Physik und Astronomie, Universit\"{a}t Potsdam, Karl-Liebknecht-Str. 24/25, 14476 Potsdam, Germany\\
$^{4}$Instituto de Astronomía, Universidad Nacional Autónoma de México, Apdo. Postal 877, C.P. 22860, Ensenada, B.C., Mexico\\
$^{5}$Institute of Astronomy and Astrophysics, Academia Sinica (ASIAA), 10617 Taipei, Taiwan\\
$^{6}$Instituto de Astronom\'{i}a y Meteorolog\'{i}a, Dpto. de F\'{i}sica, CUCEI, Universidad de Guadalajara, Av. Vallarta 2602, Arcos Vallarta, 44130 Guadalajara, Mexico
}

\pubyear{2018}

\begin{document}
\label{firstpage}
\pagerange{\pageref{firstpage}--\pageref{lastpage}}
\maketitle

\begin{abstract}
We present a multiwavelength study of the iconic Bubble Nebula
(NGC\,7635) and its ionising star BD$+$60$^{\circ}$2522. We obtained
{\it XMM-Newton} EPIC X-ray observations to search for extended X-ray
emission as in other similar wind-blown bubbles around massive
stars. We also obtained San Pedro M\'{a}rtir spectroscopic
observations with the Manchester Echelle Spectrometer to study the
dynamics of the Bubble Nebula. Although our EPIC observations are
deep, we do not detect extended X-ray emission from this wind-blown
bubble. On the other hand, BD$+$60$^{\circ}$2522 is a bright X-ray
source similar to other O stars. We used the stellar atmosphere code
PoWR to characterise BD$+$60$^{\circ}$2522 and found that this star is
a young O-type star with stellar wind capable of producing a
wind-blown bubble that in principle could be filled with hot gas. We
discussed our findings in line with recent numerical simulations
proposing that the Bubble Nebula has been formed as the result of the
fast motion of BD$+$60$^{\circ}$2522 through the medium. Our kinematic
study shows that the Bubble Nebula is composed by a series of nested
shells, some showing blister-like structures, but with little
signatures of hydrodynamical instabilities that would mix the material
producing diffuse X-ray emission as seen in other wind-blown bubbles.
Its morphology seems to be merely the result of projection effects of
these different shells.

\end{abstract}


%%%%%%%%%%%%%%%%%%%%%%%%%%%%%%%%%%%%%%%%%%%%%%%%%%

%%%%%%%%%%%%%%%%% BODY OF PAPER %%%%%%%%%%%%%%%%%%

\section{Introduction}

Diffuse X-ray emission has been found in a variety of systems: OB
associations in star forming regions, Wolf-Rayet (WR) nebulae,
planetary nebulae (PNe), and superbubbles \citep[e.g.][and references
  therein]{Ruiz2013,Gudel2008,Mernier2013,Toala2012,Townsley2014,
  RamirezBallinas2019}. This X-ray emission is the signature of the
powerful feedback from hot stars in different
environments. Theoretically, in all these systems an
adiabatically-shocked hot bubble with temperatures and electron
densities of $T=10^{7}-10^{8}$~K and
$n_\mathrm{e}\lesssim0.01$~cm$^{-3}$ is powered by strong stellar
winds ($v_{\infty}\gtrsim1000$~\kms) and, in the case of superbubbles,
the additional contribution of supernova explosions
\citep[e.g.,][]{Jaskot2011}. In contrast, detailed X-ray observations
performed with {\it Chandra} and {\it XMM-Newton} detected hot gas
with temperatures of only $T_\mathrm{X}$=[1--3]$\times$10$^{6}$~K and
electron densities much higher than expected
($n_\mathrm{e}=0.1-10$~cm$^{-3}$). This discrepancy has been
attributed to mixing processes between the hot bubble and the outer
cold nebular material due to hydrodynamical instabilities and/or
thermal conduction
\citep[e.g.,][]{Arthur2012,Dwarkadas2013,Toala2011,Weaver1977}.

\begin{figure}
\begin{center}
  \includegraphics[angle=0,width=\linewidth]{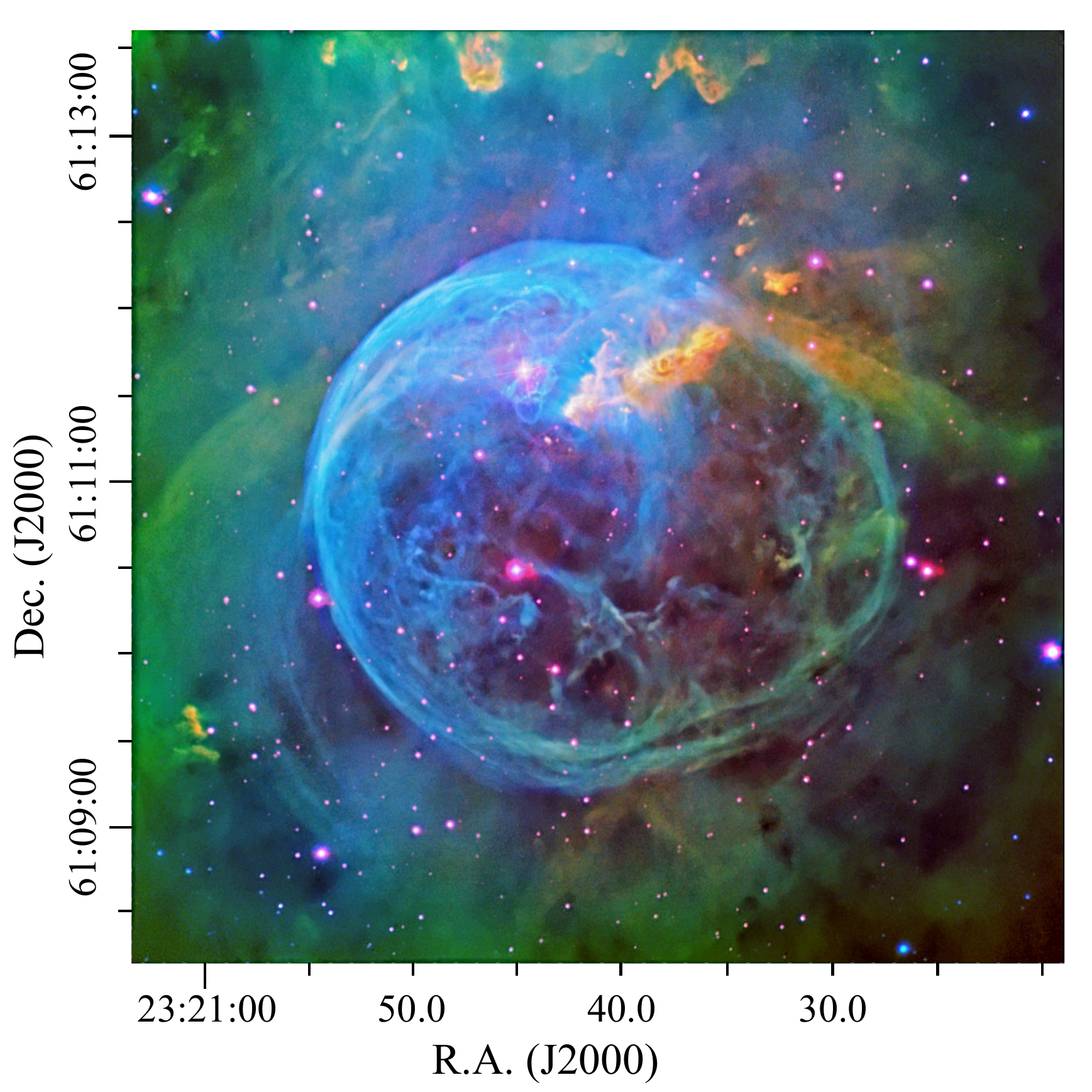}
\caption{NOT colour-composite view of the Bubble Nebula
  (a.k.a. NGC\,7635). Red, green, and blue correspond to [S\,{\sc
      ii}], H$\alpha$, and [O\,{\sc iii}].}
\end{center}
\label{fig:Bubble_Public}
\end{figure}

Among individual massive stars, diffuse X-ray emission have been
detected only in a handful of circumstellar nebulae. The most
numerous are WR nebulae \citep[see the cases of the nebulae around
  WR\,6, WR\,7, WR\,18 and
  WR\,136;][]{Toala2012,Toala2015,Toala2016a,Toala2017}. The powerful
winds from WR stars \citep[$v_{\infty} \gtrsim 1500$~km~s$^{-1}$,
  $\dot{M}\approx10^{-5}$~M$_{\odot}$~yr$^{-1}$;][]{Hamann2006}
interact with slow and dense material ejected on a previous
evolutionary stage (either a red supergiant or luminous blue variable)
creating the WR nebula. Interestingly, WR nebulae that display diffuse
X-ray emission harbour early nitrogen-rich (WNE) type stars, whilst
late WN stars do not \citep[][]{Gosset2005,Toala2013,Toala2018b}.

Although hot gas can be easily produced by the strong winds of massive
O stars, not many wind-blown bubbles (WBBs) within the H\,{\sc ii}
regions around single hot stars have been detected by X-ray
satellites. The hot bubble around the runaway O\,9.5V star
$\zeta$\,Oph is so far the only case \citep{Toala2016b}. $\zeta$\,Oph
is the closest runaway massive star at a distance of 222~pc
\citep[][]{Megier2009}. \citet{Toala2016b} argue that the soft diffuse
X-ray emission seems to be powered by hydrodynamical mixing at the
wake of the bow shock as predicted by the radiation-hydrodynamic
models presented by \citet{Mackey2015}.

It is crucial to study the feedback from single massive stars in order
to understand this effect and extrapolate to the more complex case of
OB associations. For this, we have selected a key object for a
detailed study of a WBB around an O-type star. The iconic Bubble
Nebula (a.k.a. NGC\,7635) encompasses the O-type star
BD$+$60$^{\circ}$2522 (O\,6.5\,III) and is associated to the ionised
H\,{\sc ii} region S\,162 \citep[][]{Maucherat1973}. This region has
uniquely simple morphology, which allows to disentangle the effects of
the stellar wind and the reach of the ionisation photon flux effect
(see Figure~1). Wide-field optical images as those presented by
\citet[][]{Moore2002} show that the Bubble Nebula is located within an
ionised cavity with an extension of $\sim$12\arcmin\, along the
north-south direction. More recently, the {\it Hubble Space Telescope
  (HST)} has produced an exquisite view of the Bubble
Nebula\footnote{\url{https://www.nasa.gov/sites/default/files/thumbnails/image/p1613a1r.jpg}}.
BD$+$60$^\circ$2522 is the only apparent source of ionisation and
mechanical energy. This makes the Bubble Nebula the perfect object to
study the interaction of fast stellar winds with the ionised
interstellar medium from a massive single star and the formation of a
hot bubble.

With this in mind, we have carried out a multi-wavelength study of the
Bubble Nebula in optical, UV, and X-rays. This paper is organised as
follows. In Section~2 we present our observations. Section~3 presents
the stellar atmosphere analysis of the central star of the Bubble
Nebula, BD$+$60$^{\circ}$2522. Section~4 presents our
results. Finally, the discussion and summary are presented in
Section~5 and 6, respectively.

\section{Observations}

\subsection{Optical images and spectroscopy}

We have obtained [S\,{\sc ii}], H$\alpha$, and [O\,{\sc iii}]
narrowband images of the Bubble Nebula on 2015 July 17-18 with the
Alhambra Faint Object Spectrograph and Camera (ALFOSC) at the 2.5~m
Nordic Optical Telescope (NOT) at the Observatorio del Roque de los
Muchachos, La Palma (Spain). The central wavelengths and bandpasses of
the three filters are 6725~\AA\, and 10~\AA\, for [S\,{\sc ii}],
6563~\AA\, and 33~\AA\, for H$\alpha$, and 5010~\AA\, and 43~\AA\, for
[O\,{\sc iii}], respectively. The total exposure times were 1800,
1500, and 900~s for the [S\,{\sc ii}], H$\alpha$, and [O\,{\sc iii}]
images, respectively. The averaged seeing during the observations was
$\sim$0\farcs7. The final colour-composite image of the Bubble Nebula
is presented in Figure~1.

We have also obtained cross-dispersed high-resolution Fibre-fed
Echelle Spectrograph (FIES) observations of BD$+$60$^{\circ}$2522 at
the NOT \citep{Telting2014} on 2015 July 19. The high-resolution mode
({\sc high-res}, $R=67,000$) was used to acquire a spectrum in the
3700--7300~\AA\, range without gaps in a single fixed setting. The
total exposure time was 800~s.

A set of 12 long-slit, high-resolution spectroscopic observations were
obtained at the Observatorio Astron\'omico Nacional in San Pedro
M\'artir, Mexico, using the Manchester Echelle Spectrometer (MES-SPM)
mounted on the 2.1~m telescope. Spectra were obtained on 2015 August
11--15 and another six on 2019 November 6. The slit, with a fixed
length of 5\farcs5 and width set to 150~$\mu$m ($\simeq$1\farcm9), was
placed at several positions covering different spatial features in the
nebula (see Fig. 2).  For both observation runs, we used a
2048$\times$2048 pixels E2V CCD with a pixel size of 13.5$\mu$m per
pixel with a 2$\times$2 binning corresponding to a spatial scale of
0\farcs351 per pixel. The filter centred on the [O\,{\sc iii}]
$\lambda$ 5007 emission line (with $\Delta\lambda$= 50 \AA) was used
to isolate the echelle 114th order leading to a spectral scale of
0.043~\AA~pixel${-1}$.  All the spectra were taken with exposures of
1800~s and the seeing was $\sim2''$ for all observations as estimated
from the FWHM of stars in the field. The wavelength-calibration was
performed with a ThAr arc lamp with an accuracy of
$\pm$1~km~s$^{-1}$. The spectral resolution is given by the FWHM of
the arc lamp emission lines and is estimated to be $\simeq$
12$\pm$1~km~s$^{-1}$. As we are primarily interested in the
kinematical information no flux calibration was performed. The
resultant spectra are presented in Figure~3.

All optical data (images and spectra) described in this section were
reduced using standard {\sc iraf} procedures \citep[][]{Tody1993}.

\begin{figure*}
\begin{center}
  \includegraphics[angle=0,width=\linewidth]{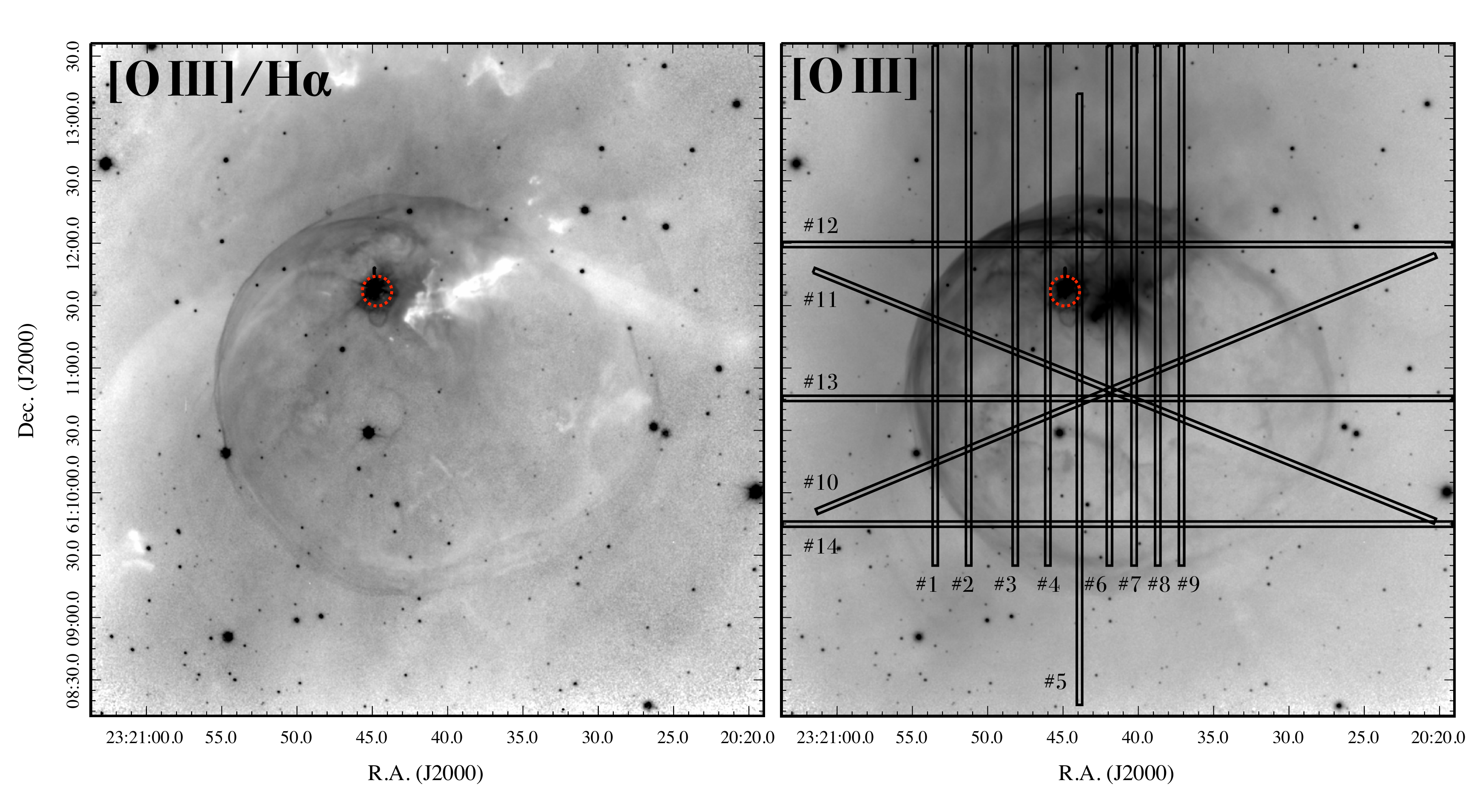}
\caption{Left: Gray-scale image of the [O\,{\sc iii}] to H$\alpha$
  ratio map of NGC\,7635. Right: [O\,{\sc iii}] image of NGC\,7635
  with the slit positions of the MES-SPM spectra overplotted. The
  position of BD$+$60$^{\circ}$2522 is shown with a (red) dashed-line
  circle. %Red labels (number$+$letter) show the position of
  %morphological features shown in Figure~10.
}
\end{center}
\label{fig:Bubble_ratio}
\end{figure*}

\begin{figure*}
\begin{center}
  \includegraphics[angle=0,width=\linewidth]{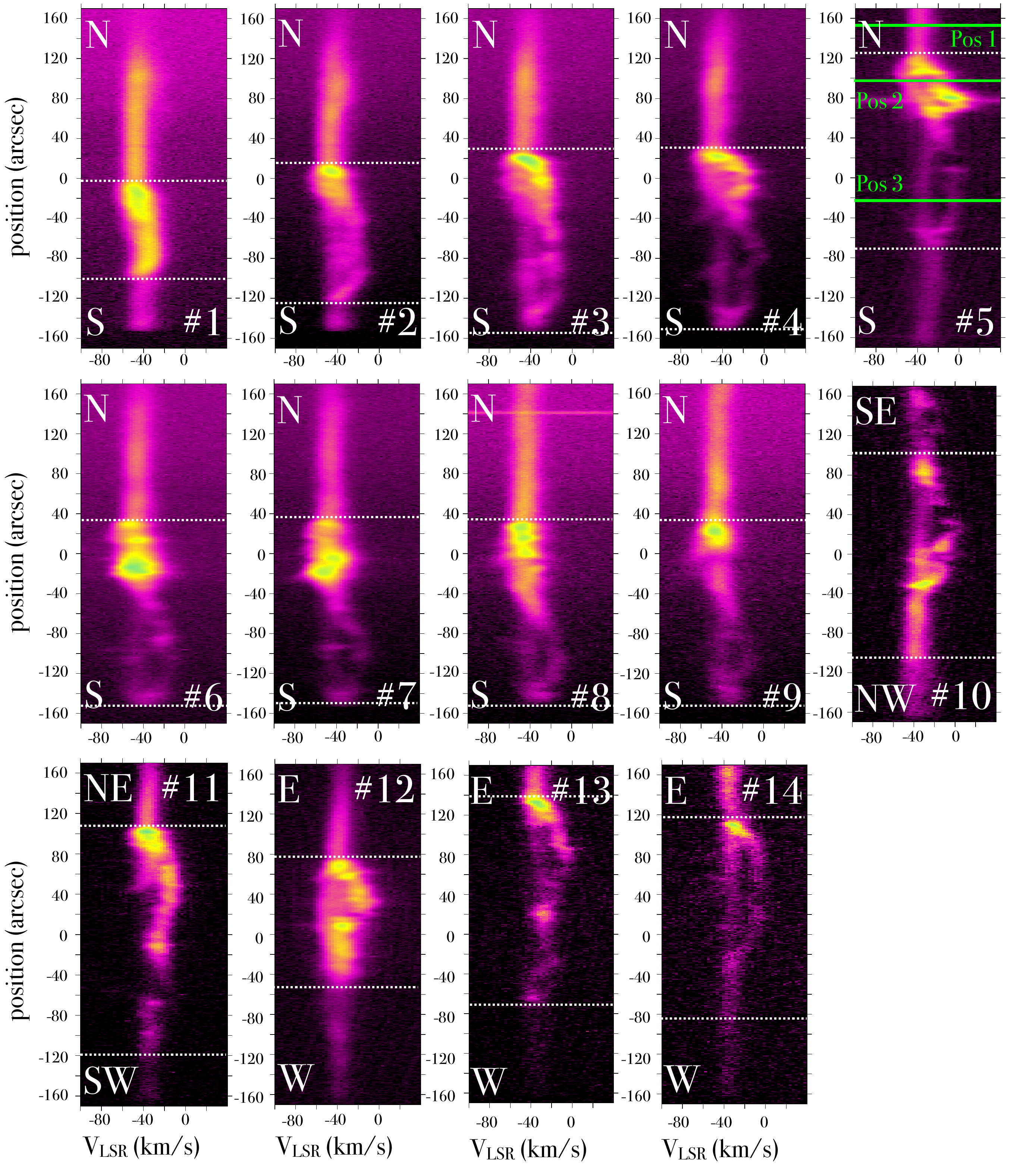}
\caption{Echellograms obtained from the SPM-MES observations of the
  Bubble Nebula. The white dashed-lines delimit the extension of the
  Bubble Nebula as seen in the [O\,{\sc iii}] narrow-band
  image. Positions 1, 2 and 3 on the spectrum from Slit\,\#5
  correspond to the extraction regions of the velocity profiles shown
  in Figure~9.}
\end{center}
\label{fig:MES}
\end{figure*}

\subsection{{\it XMM-Newton} observations}

The Bubble Nebula was observed by the European Space Agency {\it
  XMM-Newton} X-ray telescope on 2015 June 18 (Observation ID
0764640101; PI: M.A.\,Guerrero) using the European Photon Imaging
Cameras (EPIC) in Full Frame Mode with the medium optical blocking
filter. The total exposure times for the pn, MOS1, and MOS2 cameras
were 51.8, 63.5, and 63.4~ks, respectively.

In order to analyse the X-ray observations of the Bubble Nebula we
used the {\it XMM-Newton} Science Analysis Software ({\sc sas})
version 15.0 and the calibration access layer available on 2017 July
3. The Observation Data Files were reprocessed using the {\sc sas}
tasks {\it epproc} and {\it emproc} to produce the corresponding event
files.  Periods of high-background levels were removed from the
data. This has been done by creating lightcurves in the 10--12~keV
energy range binning the data over 100~s for each EPIC camera. The
background was considered to be high for count rates values higher
than 0.4, 0.18, and 0.18 for the pn, MOS1, and MOS2 cameras,
respectively.  After processing, the final effective times were
reduced to 40.5, 58.7, 58.9~ks for the pn, MOS1, and MOS2,
respectively.

To obtain a clear view of the distribution of the X-ray emission in
NGC\,7635, we followed the Snowden \& Kuntz cookbook for the analysis
of {\it XMM-Newton} EPIC observations of extended sources ({\sc
  xmm-esas}). These tasks remove the contribution from astrophysical
background, soft proton background, and solar wind charge-exchange
reactions, which contribute importantly at lower energies
($E<$1.5~keV). The {\sc esas} tasks were used to create EPIC images in
the soft (0.3--1.2~keV), medium (1.2--2.5~keV), and hard
(2.5--9.0~keV) energy bands. Individual EPIC-pn, EPIC-MOS1, and
EPIC-MOS2 images were extracted, corrected for exposure maps, and
merged together. Figure~4 presents the final exposure-map-corrected,
background-subtracted EPIC images as well as a colour-composite image
of the three bands. Each image has been adaptively smoothed using the
{\sc esas} task {\it adapt} requesting 10 counts per smoothing kernel.

\begin{figure*}
\begin{center}
\includegraphics[angle=0,width=\linewidth]{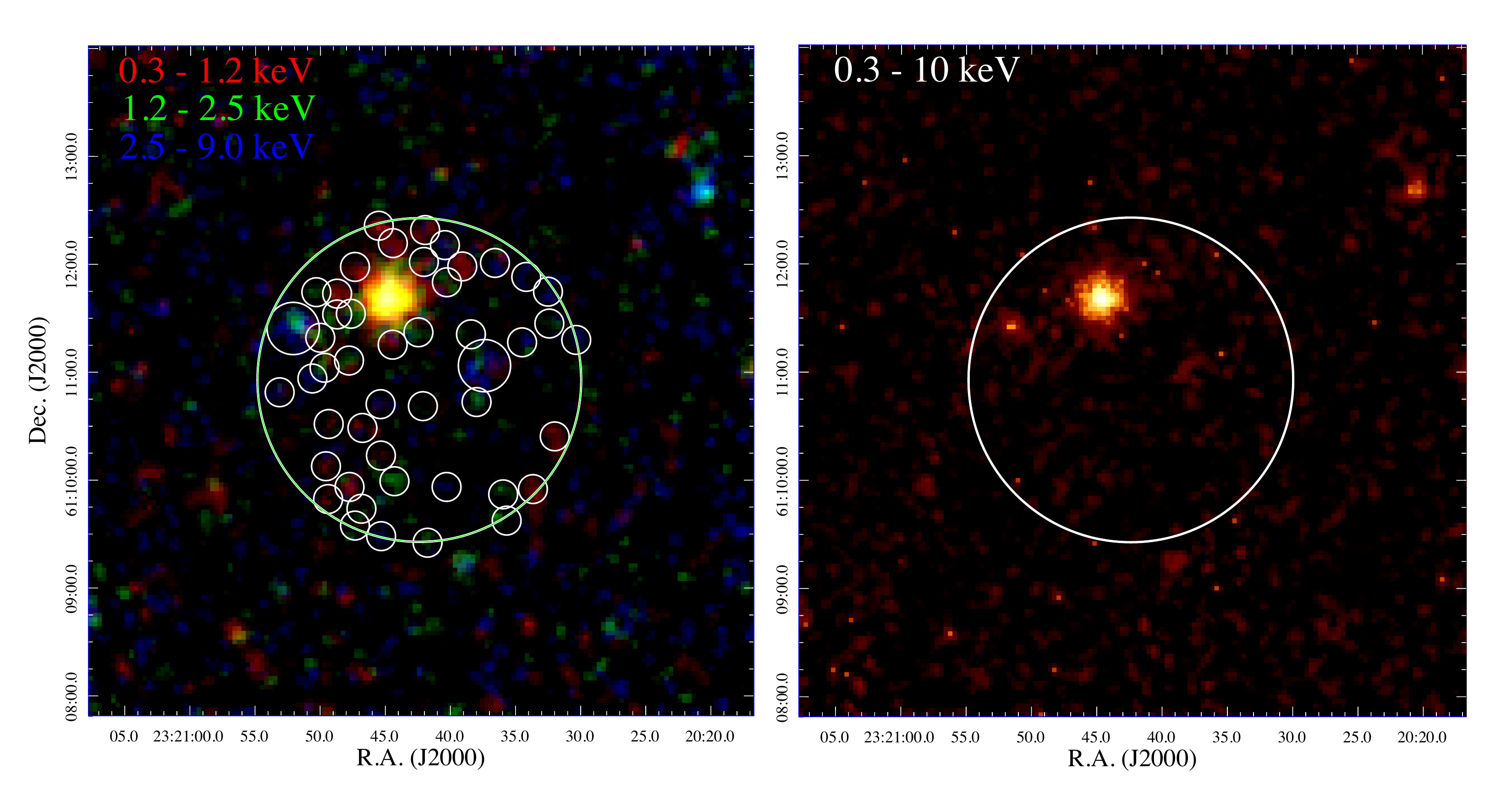}
\caption{{\it XMM-Newton} EPIC (pn$+$MOS1$+$MOS2) exposure-corrected,
  background-subtracted images in the field of view of
  NGC\,7635. Different bands are labelled on the upper-left corner on
  each panel. The left panel presents a colour-composite image using
  the three X-ray bands while the right panel shows an image with the
  complete X-ray band (0.3--10~keV). The large circular aperture in
  both panels encompasses the optical image of the nebula with a
  radius of 1\farcm7. BD$+$60$^{\circ}$2522 is the brightest source of
  X-ray emission. Point sources have not been excised from these
  images and their positions are shown by the smaller circles in the
  left panel.}
\end{center}
\label{fig:bubble_all}
\end{figure*}

\begin{figure*}
\begin{center}
\includegraphics[angle=0,width=0.7\linewidth]{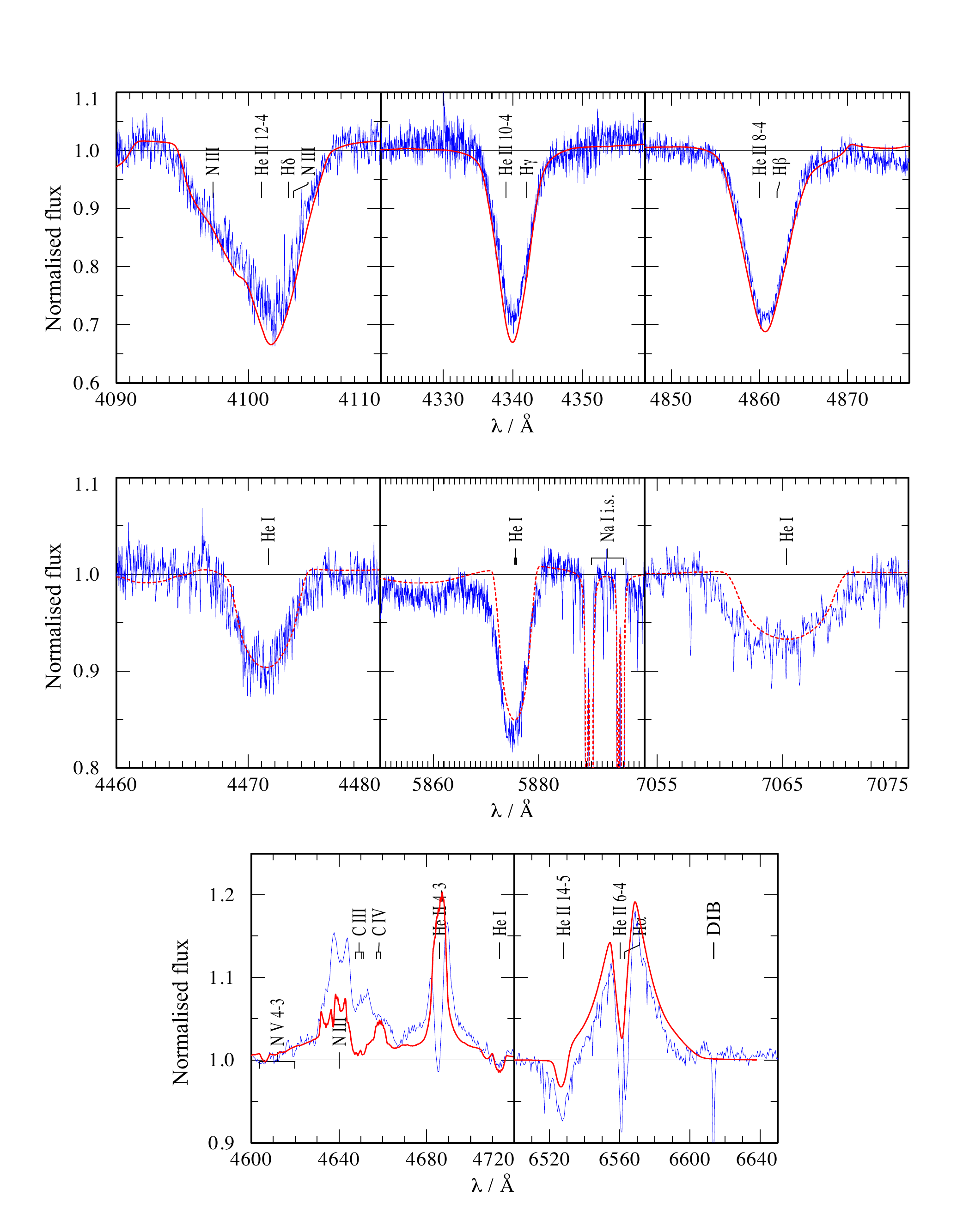}
\caption{Optical lines extracted from the FIES spectrum of
  BD$+$60$^{\circ}$2522. Normalised observation (blue) vs. synthetic
  spectrum of our best fitting model (red dashed). The absorption lines
  are rotationally broadened with $v \sin i=$178~km~s$^{-1}$. The model
  spectrum was convolved with a Gaussian with 2~\AA\, FWHM to match
  the resolution of the FIES observations, inferred from the
  interstellar Na\,{\sc i} doublet.}
\end{center}
\label{fig:optical_spec}
\end{figure*}

\begin{figure*}
\begin{center}
  \includegraphics[angle=0,width=0.7\linewidth]{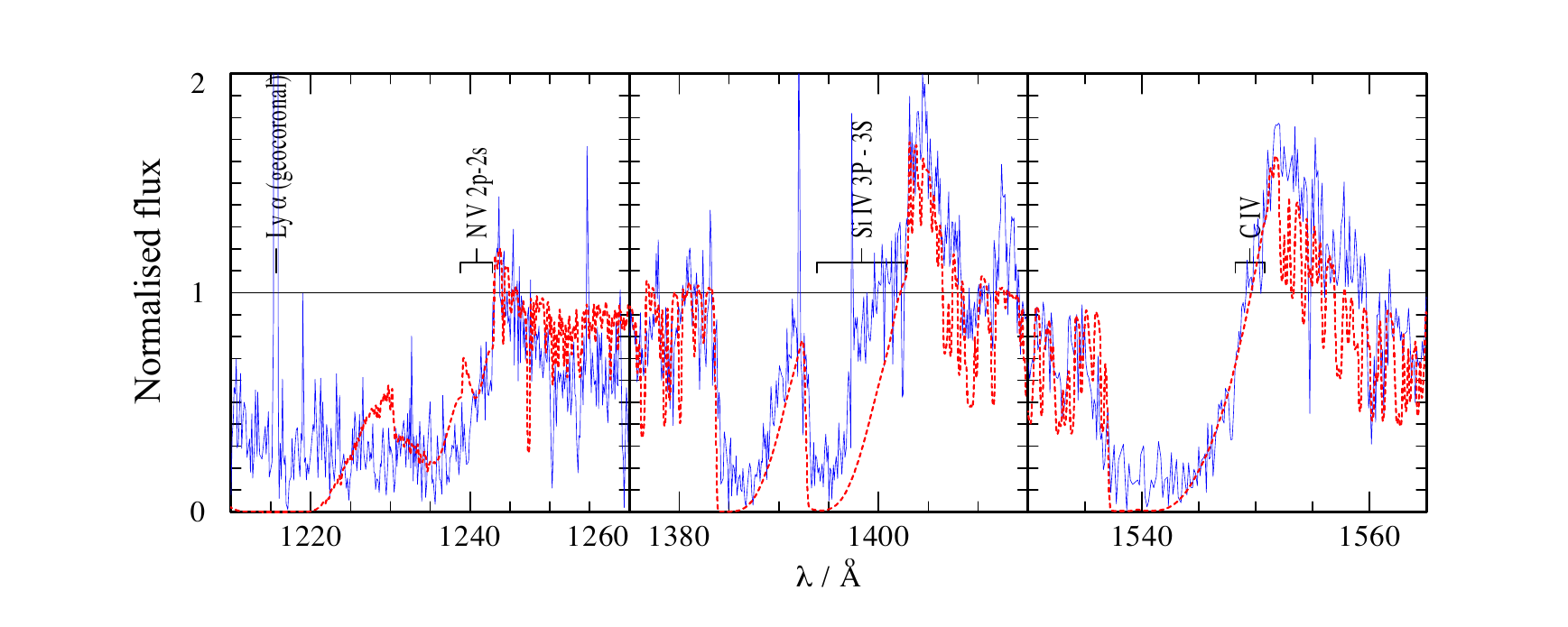}
\caption{P Cyg profiles detected in {\it IUE} UV spectrum of
  BD$+$60$^{\circ}$2522. Normalised observation (blue) vs. synthetic
  spectrum of our best fitting model (red dashed). The P Cyg profiles
  imply a terminal wind velocity of $v_{\infty}=2000$~km~s$^{-1}$.}
\end{center}
\label{fig:optical_UV}
\end{figure*}

\begin{figure*}
\begin{center}
\includegraphics[angle=0,width=1.\linewidth]{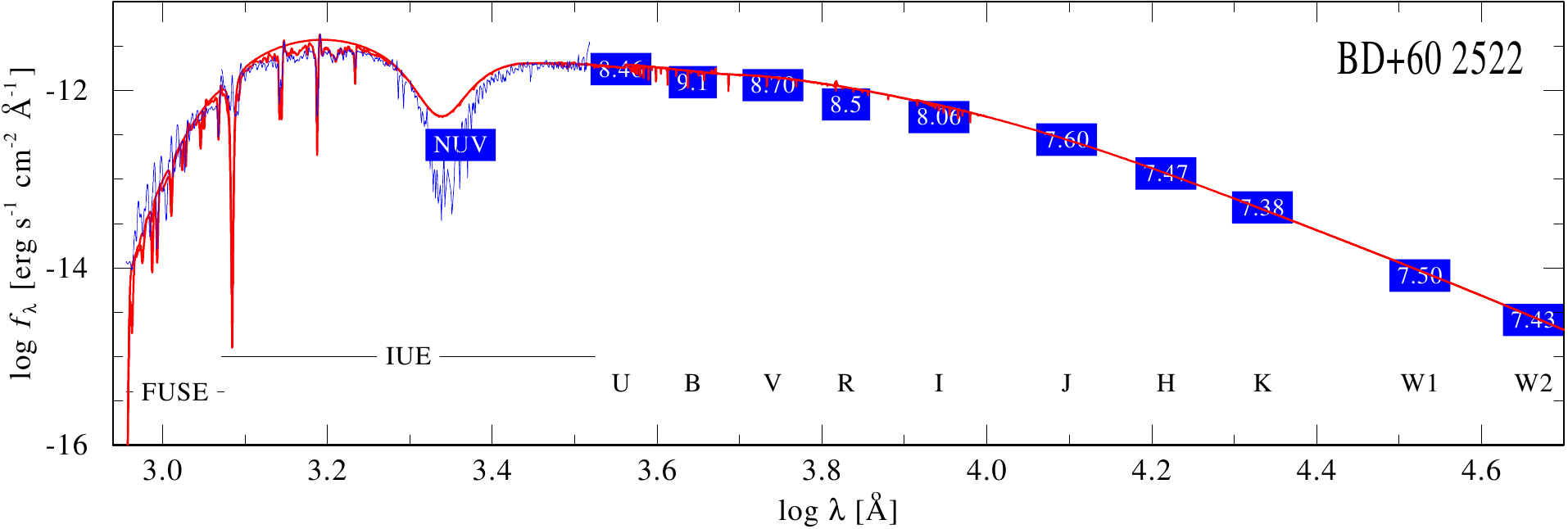}
\caption{Spectral Energy Distribution of BD$+$60$^{\circ}$2522 from
  the UV to the IR range (in blue). Blue squares are photometric
  measurements in the indicated bands. The SED obtained for our
  best-fitting model is shown in red.}
\end{center}
\label{fig:SED}
\end{figure*}

\subsection{Complementary Archival Observations}

In order to model the atmosphere of BD$+$60$^{\circ}$2522 we
downloaded {\it Far Ultraviolet Spectroscopic Explorer} ({\it FUSE})
and {\it International Ultraviolet Explorer} ({\it IUE}) observations.
The data from these observations have been retrieved from MAST, the
Multimission Archive at the Space Telescope Science
Institute\footnote{STScI is operated by the Association of
  Universities for Research in Astronomy, Inc., under NASA contract
  NAS5-26555.}. The {\it FUSE} observations correspond to the IDs
d1130101000 (PI: P.\,Dufour) and u1046701000 (PI: W.P.\,Blair)
obtained on 2003 August 4 and 2006 November 2 with total exposure
times of 35~ks and 16~ks, respectively. The {\it FUSE} observations
cover a spectral range between 916--1190~\AA.

The {\it IUE} observations in the spectral range 1149-1978~\AA\,
correspond to the Obs.\,ID swp08840 and have been taken with the large
aperture at high dispersion with a total exposure time of 10~ks.

In total there are 13 {\it IUE} and four {\it FUSE} observations in
the archives with different spectral settings or from different
epochs. The data scarcity precludes a study of UV line variability.
\citet{Rauw2003} reported the variability of some optical emission
lines in BD$+$60$^{\circ}$2522, e.g., He\,{\sc ii}\,4686, but it is
small if compared to the Of?p star CPD$-$28${^\circ}$2561
\citep[e.g.][]{hubrig2015}.

\section{Analysis of the stellar wind from BD$+$60$^{\circ}$2522}

We analysed the optical FIES and UV {\it FUSE} and {\it IUE} spectra
of BD$+$60$^{\circ}$2522 using the most recent version of the Potsdam
Wolf-Rayet (PoWR) model
atmosphere\footnote{\url{http://www.astro.physik.uni-potsdam.de/PoWR}}.
The PoWR code solves the NLTE radiative transfer problem in a
spherical expanding atmosphere simultaneously with the statistical
equilibrium equations and accounts at the same time for energy
conservation. Iron-group line blanketing is treated by means of the
superlevel approach \citep{Grafener2002}, and wind clumping is taken
into account in first-order approximation \citep{Hamann2004}. We do
not calculate hydrodynamically consistent models, but assume a
velocity field following a $\beta$-law with $\beta$ = 0.8 as used
e.g.\ in \citet{shenar2015} for $\delta$\,Ori, which gives a
consistent fit for most of the the wind lines, i.e., a depth-depended
clumping with a maximum value of $D=20$ in the outer wind. Our
computations applied here include complex atomic models for hydrogen,
helium, carbon, nitrogen, silicon, phosphorus, and the iron-group
elements.

The blue edge of the P Cygni profiles were used to estimate a terminal
wind velocity $v_{\infty}$=2000$\pm$100~km~s$^{-1}$, in agreement with
\citet{Prinja1990}.  Additional broadening due to depth dependent
microturbulence with $v_\text{D}=20\,$km~s$^{-1}$ in the photosphere
up to $v_\text{D}=100\,$km~s$^{-1}$ in the outer wind was taken into
account for the emergent spectrum, allowing an adequate fit of the
width of the Si\,{\sc iv} and the C\,{\sc iv} resonance lines (see
Fig.~6). A better fit to the blue edge of the P Cygni trough of
C\,{\sc iv} can be achieved by a slightly higher terminal velocity
($2100$\,km~s$^{-1}$) or a larger value of the microturbulence
($200$\,km~s$^{-1}$ in the outer wind). A constant value of
$v_\text{D}=50\,$km~s$^{-1}$ was used during the calculation of the
population numbers and is consistent with the observed strength of the
He\,{\sc ii} 4-3 line, which is very sensitive to the value of
$v_\text{D}$. Other authors give higher values for $v_\infty$ but
without taking microturbulence into account.

We calculated the quasi-hydrostatic part of the atmosphere
consistently according to \cite{sander2015} and took pressure
broadening of the spectral lines in the formal integral into
account. The line wings of the Balmer lines were used to determine
$\log g = 3.5$. We also applied rotational broadening to the formal
integral as described in \citet{shenar2014} with $R_\text{corot} =
R_\star$ and $v \sin i = 200\,$km~s$^{-1}$, which gives a better fit
to the photospheric absorption lines (e.g.\ He\,{\sc
  i}\,$\lambda$\,4472\,\AA, O\,{\sc iii}\,$\lambda$\,5600\,\AA) than
the value of $v \sin i = 178\,$km~s$^{-1}$ \citep{Howarth1997}.

Based on spectra and photometry from near-UV to the near-IR {\it WISE}
band at 12~$\mu$m we computed a colour-excess $E(B-V)$=0.68~mag
following the extinction law by \citet{Cardelli1989} with
$R_V$=3.1. As $A_V = R_V * E(B-V)$, the extinction in the $V$ band was
estimated to be 2.1~mag. An hydrogen column density of
$N_\text{H}=2.7\times 10^{21}\,\text{cm}^{-2}$ was estimated using the
relation given by \citet{Groenewegen1989}. The synthetic spectrum was
then corrected for interstellar extinction due to dust by the
reddening law of \citet{Seaton1979} for the UV and optical, as well as
for interstellar line absorption for the Lyman series in the UV
range. Finally, we diluted the synthetic SED by the distance. The
later was taken to be $d$ = 2.5~kpc favoured by \citet{Moore2002},
which is also consistent with the {\it Gaia} estimate of
2.5$\pm$0.2~kpc from \citep[][]{Gaia2016,Gaia2018,BJ2018}. A
comparison of our resultant "corrected" synthetic SED to the
observations is presented in Figure~7.

The effective temperature $T_\mathrm{eff} = 35\,$kK, which we define
at a radius of $\tau_\text{Rosseland}=20$, was derived from the
strength of the He\,{\sc i} lines (see Fig.~5).  The value of
$T_\mathrm{eff}$ is very well constrained; for $T_\mathrm{eff}=34\,$kK
the He\,{\sc i} lines are already much stronger than observed, while
for $T_\mathrm{eff}=36\,$kK the He\,{\sc i} lines appear to weak.

The mass-loss rate of $\log \dot{M} = -5.9\,M_\odot\,\text{yr}^{-1}$
was determined with help of the H$\alpha$ emission line, but is also
consistent with the UV resonance and optical emission lines (see
Fig.~5 middle panel). However, we failed to reproduce the C\,{\sc
  iii}\,$\lambda\lambda$\,4647.4 4650.3 4651.1-multiplet in emission
as observed, as our models show these lines only in absorption. Note
that this feature is stronger in our observation than in those by
\citet{Rauw2003}.  As the model reproduces the C\,{\sc
  iii}\,$\lambda$\,5696 line sufficiently well, it does not seem to be
a problem of the C\,{\sc iii}/C\,{\sc iv} ionisation balance in
general.

The observed P Cygni line profile of the N\,{\sc v} resonance doublet
can only be reproduced by taking Auger ionisation due to X-rays in the
wind into account. We adopted an X-ray luminosity of about
$L_\text{X}/L_\text{bol} \approx -7$ and a plasma temperature derived
in the X-ray analysis (see next section). We account only for the
free-free emission of these hot electrons. The X-rays then ionise the
N\,{\sc iii} of the ``warm'' wind to N\,{\sc v}.

%%%%%%%

The parameters of our best-fit PoWR model to BD$+$60$^{\circ}2522 $are
presented in Table~1. The comparison between normalised spectral lines
as detected by FIES and UV observations with our best-fit model are
presented in Figures~5 and 6, respectively.

\begin{table}
\caption{Parameters of the best-fit PoWR model of BD$+$60$^{\circ}$2522.}
\begin{tabular}{lcl}
\hline
Parameter & Value & Comment\\
\hline
$E({B-V})$ [mag]                                   & 0.68$\pm$0.02   & fitted \\
$T_\text{eff}$ [kK]                                 &  35$\pm$0.5     &  fitted \\
$d$ [kpc]                                           &  2.5$\pm$0.2   & from \citet{BJ2018} \\
$\log(L_\star/$L$_\odot)$                            &  5.4$\pm$0.1    & fitted  \\
$R_\star$ [R$_\odot$]                                &  15.3$\pm$0.4   & from $L_{\star}$ and $T_\mathrm{eff}$\\
$D$ (clumping factor)                               &  20             & depth-dependent \\
$\log(\dot{M}/\mathrm{M}_\odot\,\text{yr}^{-1})$     & $-$5.9$\pm$0.1  & from H$\alpha$  \\
$v_{\infty}$ [km\,s$^{-1}$]                          & 2000$\pm$100   & fitted \\
log\,$g$ [cm~s$^{-2}$]                                & 3.5$\pm$0.1    & fitted \\
$M_\star$ [M$_\odot$]                                & 27$\pm$7       & from $d$ and log\,$g$ \\
\hline
\multicolumn{3}{c}{Chemical abundances (mass fraction)}\\
\hline
H  & 0.74               & solar \\
He & 0.25               & solar \\
C  & (5.4$^{+6.6}_{-3.0}$)$\times10^{-4}$ & 0.25$\times$solar\\
N  & (2.8$^{+9.7}_{-1.4})\times10^{-3}$ & 4$\times$solar\\
O  & (1.2$^{+0.3}_{-0.3}$)$\times10^{-3}$ & 2$\times$solar\\
Si & 6.7$\times10^{-4}$ & solar \\
P  & 5.8$\times10^{-6}$ & solar \\
S  & 3.1$\times 10^{-4}$ & solar \\
Fe-group & 1.3$\times10^{-3}$ & solar\\
\hline
\end{tabular}
\end{table}

\section{Results}

\subsection{X-rays from NGC7635 and BD$+$60$^{\circ}$2522}

\begin{figure}
\begin{center}
\includegraphics[angle=0,width=1.\linewidth]{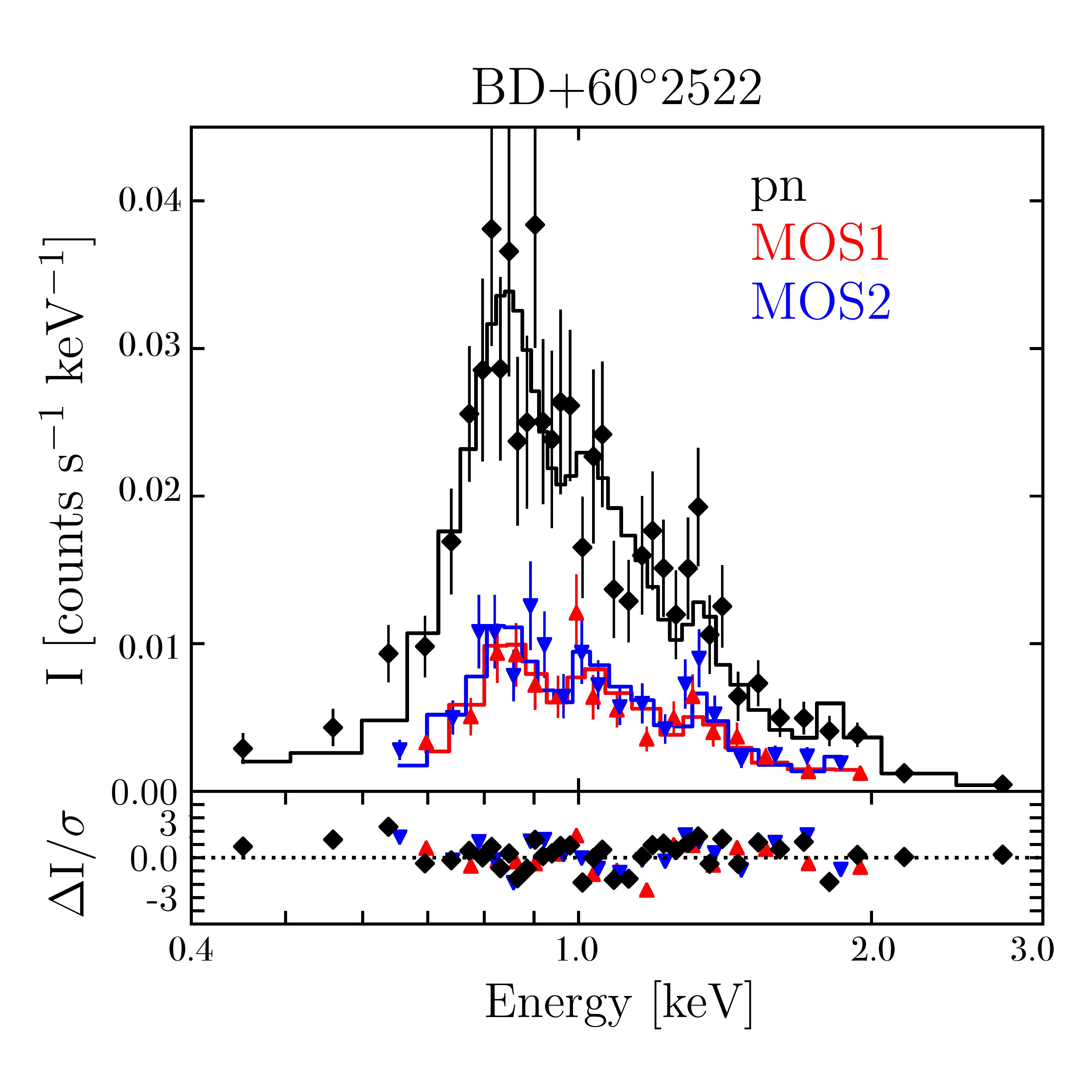}
\caption{{\it XMM-Newton} background-subtracted spectra of the central
  star of the Bubble Nebula (BD$+$60$^{\circ}$2522). Solid lines
  represent the best-fit models.}
\end{center}
\label{fig:EPIC_spec}
\end{figure}

Figure~4 presents X-ray images as described in Secion~2.3. Although
our {\it XMM-Newton} observations are deep, there is no clear evidence
of diffuse X-ray emission filling NGC\,7635. Most of the detected
emission within the Bubble Nebula can be explained to the contribution
of point sources along the line of sight. Some of them even have
optical and IR counterparts. The rest have been identified by
the pipeline of identification of point-sources. Thus, we conclude
that no diffuse X-ray emission is detected within NGC\,7635.

To calculate an upper limit to the diffuse X-ray emission, we
extracted the spectrum from the large circular aperture defined in
Figure~4. This has been done by excising regions that contain all
identified point sources, regardless of the the band they were
identified, in order to reduce any possible contribution. We derived a
3$\sigma$ upper limit to the EPIC-pn camera in the 0.3--5.0~keV of
1.5$\times$10$^{-3}$~counts~s$^{-1}$ after correcting for the area of
the excised regions. We used the PIMMS
webpage\footnote{\url{http://cxc.harvard.edu/toolkit/pimms.jsp}} to
estimate the flux upper limit of the observation adopting an
$E(B-V)$=0.73~mag. For discussion (see Section~5), we assumed an
optically thin model plasma {\it apec} model with solar abundances for
two different plasma temperatures: i) a soft temperature
$kT_\mathrm{soft}=$0.22~keV ($T_\mathrm{soft}$=2.5$\times$10$^{6}$~K)
and ii) a hard temperature of $kT_\mathrm{hard}$=2.16~keV
($T_\mathrm{hard}$=2.5$\times$10$^{7}$~K). The estimated upper limits
of the unabsorbed flux for the soft and hard temperatures are
$F_\mathrm{soft,X}<$1$\times$10$^{-14}$~erg~cm$^{-2}$~s$^{-1}$ and
$F_\mathrm{hard,X}<$6$\times$10$^{-15}$~erg~cm$^{-2}$~s$^{-1}$. These
fluxes correspond to X-ray luminosity upper limits of
$L_\mathrm{soft,X}<9\times10^{30}$~erg~s$^{-1}$ and
$L_\mathrm{hard,X}<$5$\times10^{30}$~erg~s$^{-1}$. According to PIMMS
these correspond to normalisation parameters\footnote{The
  normalisation parameter is defined as $A=10^{-14} \int n_\mathrm{e}
  n_\mathrm{H} dV/4 \pi d^2$} of
$A_\mathrm{soft}$=8.0$\times$10$^{-6}$~cm$^{-5}$ and
$A_\mathrm{hard}$=4.3$\times$10$^{-6}$~cm$^{-5}$.

On the other hand, Figure~4 clearly shows that X-ray emission is
unambiguously detected from the central star of NGC\,7635,
BD$+$60$^{\circ}$2522. In order to study the physical properties of
this X-ray emission, we have extracted background-subtracted spectra
from the three EPIC cameras. These were produced using a circular
aperture with 20$\arcsec$ in radius. Figure~8 presents the resultant
background-subtracted EPIC-pn and MOS spectra of
BD$+$60$^{\circ}$2522. Most of the emission is detected in the
0.4--3.0~keV energy range, peaking at 0.8--0.9~keV, signature of the
Ne lines and the Fe-complex.

We have modelled the X-ray emission from BD$+$60$^{\circ}$2522 using
an {\it apec} emission model. The best-fit model ($\chi^{2}$/DoF=1.15)
resulted in a main plasma temperature of $kT$=0.60~keV ($T \approx 7
\times10^{6}$~K). The absorbed flux in the 0.4--3~keV energy range is
$f_\mathrm{X}$=4.1$\times$10$^{-14}$~erg~cm$^{-2}$~s$^{-1}$ which
corresponds to an unabsorbed flux of
$F_\mathrm{X}$=3.6$\times$10$^{-13}$~erg~cm$^{-2}$~s$^{-1}$. Thus, the
estimated X-ray luminosity at a distance of 2.5~kpc is
$L_\mathrm{X}$=2.5$\times$10$^{32}$~erg~s$^{-1}$. Using the absolute
magnitude of $M_\mathrm{bol}$=$-$9.7~mag \citep[see table~1
  in][]{Chris1995} we estimated a bolometric luminosity of
$L_\mathrm{bol}$=6$\times$10$^{5}$~L$_{\odot}$, thus, fulfilling the
$L_\mathrm{X}\approx$10$^{-7}$~$L_\mathrm{bol}$ relationship of O-type
stars \citep[][and references
  therein]{Pallavicini1981,Nebot2018}. This model is shown in Figure~8
in comparison with the EPIC spectra.

We also searched for X-ray variability from BD$+$60$^{\circ}$2522. For
this, we created lightcurves in different energy ranges. %We show in
%Fig.~8 light curves extracted from the three EPIC instruments from the
%complete energy range 0.3--3.0~keV.
There is no apparent variation in the X-ray light curve in the
complete energy range. Different bands have been also analysed, but a
similar result are found. %To illustrate this, we also show in Fig.~8 -
%bottom panel the EPIC-pn light curves extracted for the 0.3--1.0~keV
%and 0.7--2.0~keV energy bands.
We note that \citet{Rauw2003} reported
that BD$+$60$^{\circ}$2522 exhibits a variability of 2--3 days which
is several times larger than the exposure time of our EPIC
observations.

\subsection{The kinematics of NGC\,7635}

Our SPM-MES echelle observations presented in Figure~3 reveal the
kinematics of the Bubble Nebula. The complex velocity structure
clearly departs from the spherical case \citep[see][and references
  therein]{Chris1995}. Although the general extension of the Bubble
Nebula can be traced in the PV diagrams, there is noticeable
kinematical substructures. Several spectra presented in Figure~3
exhibit a double bubble morphology. A small bubble-like structure
surrounding BD$+$60$^{\circ}$2522 and a more extended secondary
structure well-resolved in velocity towards the south.

In order to unveil the detailed structure of the double bubble
morphology, we extracted velocity profiles in three different
positions from Slit\,\#5. Position~1 (Pos~1) corresponds to the region
outside the Nubble Nebula towards the north, position~2 (Pos~2) was
placed at the bubble-structure around the central star and Position~3
(Pos~3) corresponds to the larger bubble-structure (see Fig.~3). The
corresponding velocity profiles extracted from the three positions in
Slit\,\#5 are presented in Figure~9. The velocity profile
extracted from Pos~1 corresponds to the V$_\mathrm{LSR}$ of the
ionised complex to the north of the Bubble Nebula. This presents a
single peak profile at V$_\mathrm{LSR}$=$-$36~km~s$^{-1}$
(V$_\mathrm{hel}$=$-$44.4~km~s$^{-1}$). The expansion velocity of the
bubble around the central star extracted from Pos~2 is
$\sim$15~km~s$^{-1}$ and is centred at
V$_\mathrm{LSR}$=$-$32~km~s$^{-1}$
(V$_\mathrm{hel}$=$-$39.8~km~s$^{-1}$). The most extended bubble
structure towards the south probed by the line profile at Pos\,3 has
an estimated expansion velocity of 14.5~km~s$^{-1}$ and it is centred
at V$_\mathrm{LSR}$=$-$19~km~s$^{-1}$
(V$_\mathrm{hel}$=$-$26.8~km~s$^{-1}$). We note that there is emission
detected at V$_\mathrm{LSR}\approx-60$~km~s$^{-1}$ in the vicinity of
BD$+$60$^{\circ}$2522 corresponding to
V$_\mathrm{hel}\approx-$70~km~s$^{-1}$ which might be the reason why
most authors report such high heliocentric velocities \citep[see
  table~2 in][]{Chris1995}. We note that similar velocity profiles are
obtained if extracted from different slits. For example, a velocity
profile from Slit\,\#4, which maps the bright emission around
BD$+$60$^{\circ}$2522, has a similar velocity profile as that
illustrated by Pos~2 in Figure~9.

The general velocity structure of the Bubble Nebula is well described
by our SPM-MES observations. These observations show that the
structure close to the progenitor star is coherent and has a well
defined shape. The secondary diffuse bubble-like structure towards the
south-west region is diffuse, which makes is difficult to trace its
limits in the velocity profiles due to confusion with the outer
ionised region (see, e.g., Slit\,\#11 in Fig.~3). This double-bubble
morphology is easily traceable in the velocity profiles from
Slits\,\#1 to \#9, which are oriented in the N-S direction. For
further illustration, we extracted intensity profiles from different
SPM-MES slits. Figure~10 shows the results for Slits\,\#5 and
Slit\,\#11 in comparison with profiles extracted from the NOT [O\,{\sc
    iii}] narrow-band image. The extension of the bubble-like
structures as inferred from Figure~3 are illustrated with
arrows. Bubble~1 shows the structure around BD$+$60$^{\circ}$2522,
whilst Bubble~2 corresponds to the larger cavity centred at
V$_\mathrm{LSR}=-$19~km~s$^{-1}$. From the profile extracted from
Slit\,\#11 we can identify a large cavity (labelled as Bubble~3) that
might be part of (or related to) Bubble~2. We note that similarly to
the nebular narrow-band images presented in Figure~1 and 2, the MES
observations show that the edge of the bubble is not sharp towards the
south, west and south-west regions. As a matter of fact, the MES
observations imply that the large bubble towards the south has a
disrupted bubble morphology.

\begin{figure}
\begin{center}
\includegraphics[angle=0,width=1.\linewidth]{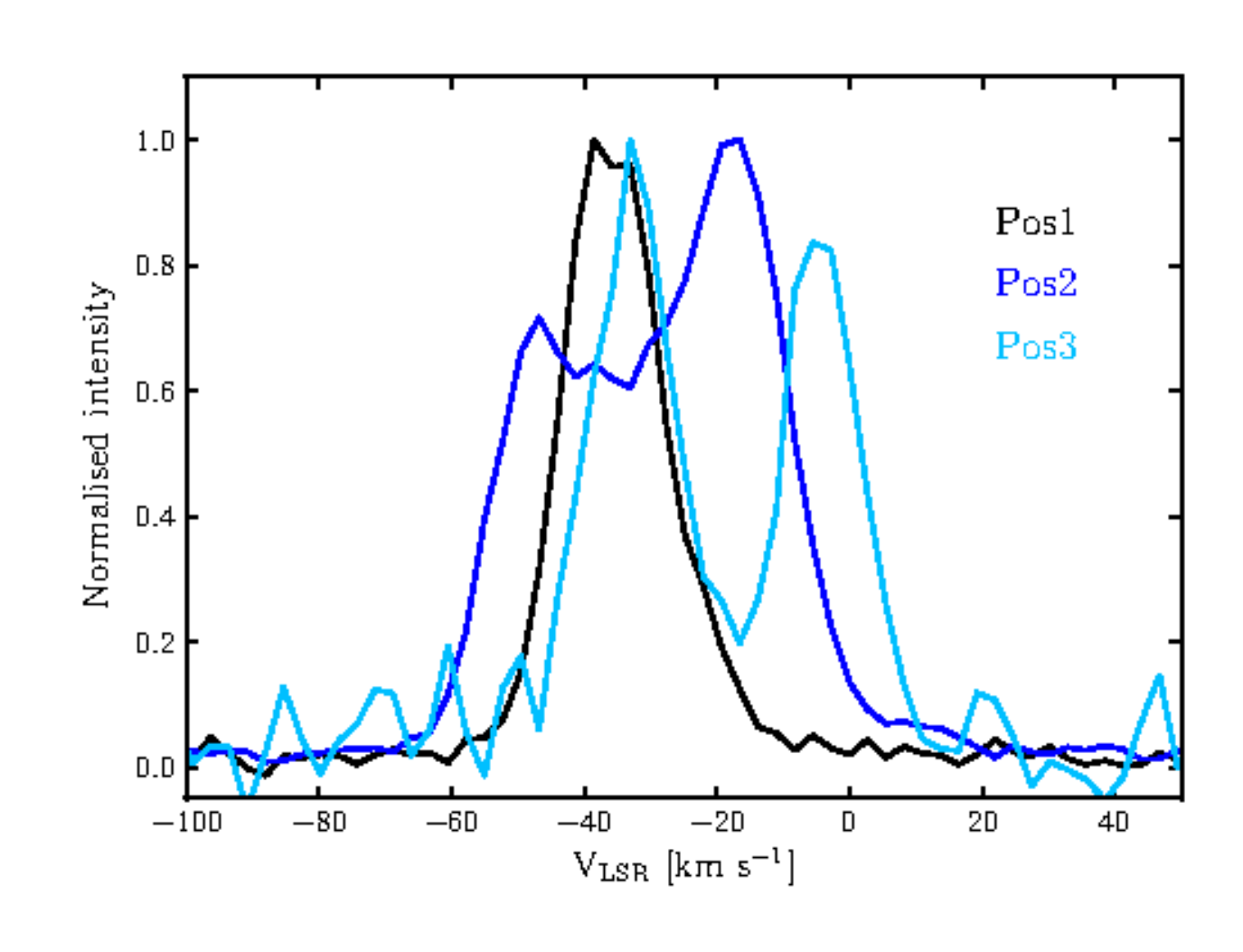}
\caption{Normalised velocity profiles extracted from different
  features from Slit\#\,1. The profiles were extracted from
  positions 1, 2 and 3 shown in Figure~3 top panel.}
\end{center}
\label{fig:vel_profiles}
\end{figure}

\begin{figure}
\begin{center}
  \includegraphics[angle=0,width=1.\linewidth]{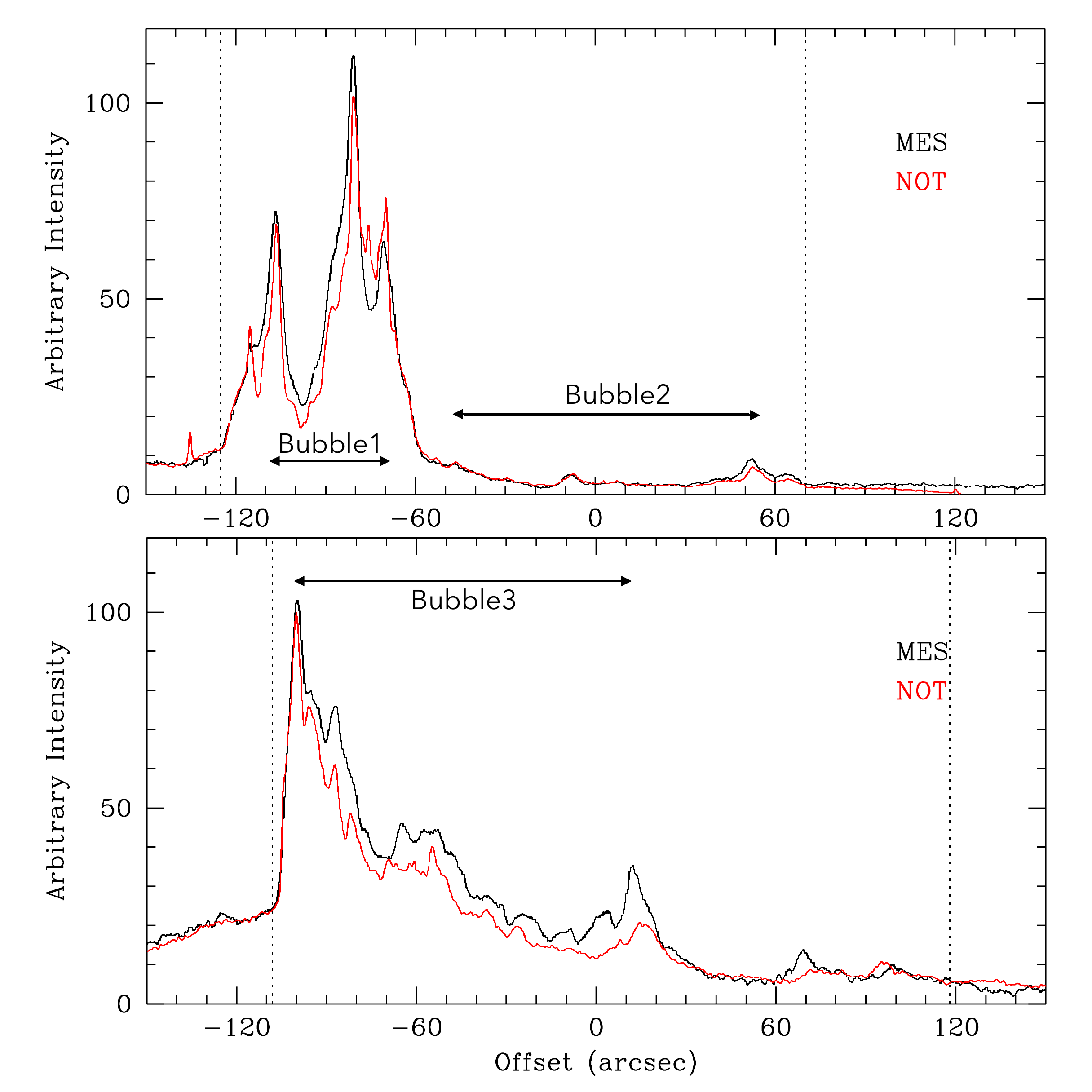}
\caption{Intensity profiles extracted from the SPM-MES observations
  (black lines) and NOT [O\,{\sc iii}] narrow-band image (red
  line). The top and bottom panel show the intensity profiles
  extracted from regions corresponding to Slit\,\#1 and Slit\,\#3,
  respecively. The vertical dotted lines delimit the extension of the
  Bubble Nebula.}
\end{center}
\label{fig:profiles}
\end{figure}

\section{Discussion}

Our detailed model of BD$+$60$^{\circ}$2522 indicates that it is a
normal O-type star with abundances close to Solar.  Its X-ray spectra
also resemble those of other O-type stars.  Although optical
variability has been reported in previous works, the duration of the
X-ray observations presented here are not long enough to make a
reliable assessment of variability in this band.

It has been widely accepted that the Bubble Nebula has been carved by
the stellar wind from BD$+$60$^{\circ}$2522.  Indeed its powerful
stellar wind ($v_{\infty}$=2000~km~s$^{-1}$,
$\dot{M}$=1.3$\times$10$^{-6}$~M$_{\odot}$~yr$^{-1}$; see section~3)
should have easily produced a bubble filled with adiabatically-shocked
wind material.  Theoretically, the stellar wind-ISM interaction
produces a hot bubble with temperature in excess to 10$^{7}$~K, but
very low electron densities, of the order of
$n_\mathrm{e}$=10$^{-3}$~cm$^{-3}$.  Most WBB detected in X-rays
exhibit plasma temperatures at least an order of magnitude below
theoretical expectations and electron densities of the order of
1--10~cm$^{-3}$ (see Section~1).  In such systems, hydrodynamical
instabilities created at the edge of the adiabatically-shocked region
in contact with the nebular material are capable of reducing the
temperature of the bubble while increasing its density and, thus,
increasing its X-ray emissivity \citep[e.g.,][]{Toala2018}.  The lack
of X-ray emission from the Bubble Nebula questions this is actually
the case.

Alternatively, \citet{Green2019} proposed that the Bubble Nebula is a
bow shock around a runaway star, following the reported high proper
motions of BD$+$60$^{\circ}$2522 (28$\pm$3~km~s$^{-1}$) from the {\it
  Gaia} data release.  As these authors propose, the Bubble Nebula
would not be a simple WBB, but rather ``a favourably oriented dense
bow shock'' located at the north of the star. These authors presented
detailed 2D hydrodynamical simulations to interpret its optical and
near-IR morphology and to predict the spatial and spectral properties
of the X-ray emission. Their best-fit model, for an ISM with constant
density of $n_{0} \approx$100~cm$^{-3}$ and a stellar velocity of 20
km~s$^{-1}$, is able to reproduce accurately the morphology of the
Bubble Nebula observed in \emph{HST} H$\alpha$ and \emph{Spitzer}
24~$\mu$m images by assuming that BD$+$60$^{\circ}$2522 moves along a
direction tilted by 60$^\circ$ with respect of the plane of the sky.
These estimates can be refined using the radial velocity of
BD$+$60$^{\circ}$2522 of $-$26$\pm$1~km~s$^{-1}$ derived from our
high-resolution FIES spectrum and the velocity on the plane of the sky
of $\simeq$25 km~s$^{-1}$ derived from our re-analysis of {\it Gaia}
data using a distance of 2.5 kpc.  Accordingly, BD$+$60$^{\circ}$2522
is found to move along an angle of 46$^{\circ}$ with respect to the
plane of the sky with a space velocity of 36 km~s$^{-1}$.  The angle
is close to that of 56$^{\circ}$ estimated by \citet{Green2019}, but
the space velocity is notably larger than their space velocity of 20
km~s$^{-1}$.

In \citet{Green2019}'s models, the hot bubble becomes unstable at the
wake of the bow shock producing hydrodynamical instabilities that are
then a source of mass, mixing material into the hot bubble to produce
optimal conditions for soft X-ray emission.  \citet{Green2019}
estimated that such process would produce an unabsorbed X-ray
luminosity for the soft X-ray {\it XMM-Newton} band (0.3--2.0~keV) of
$L_\mathrm{X}$=10$^{32}$--10$^{33}$~erg~s$^{-1}$ and a corresponding
estimate for the hard X-rays of 10$^{30}$--10$^{31}$~erg~s$^{-1}$.
Their estimated mean temperature of the soft and hard X-ray emission
are 0.22~keV and 2.2~keV, respectively. However, our observations did
not detect diffuse X-ray emission and the upper limits derived in
Section~4.1 are at least two orders of magnitude below these
theoretical predictions.

We also checked {\it FUSE} spectra looking for absorption or emission
of the [O\,{\sc vi}]~$\lambda\lambda$1032,1037 in the line of sight of
BD$+$60$^{\circ}$2522.  Along with N\,{\sc v} and C\,{\sc iv}, this
feature has been correlated with the presence of a mixing region
between the hot bubble and the nebular material and can be related to
the presence of X-rays \citep[][]{Gruendl2004,Ruiz2013,Fang2016}.  We
did not found any hint of emission nor absorption of this line.

The lack of diffuse X-ray emission from the Bubble Nebula might
suggest that mixing has not been very efficient.  This is supported by
the lack of strong indications of instabilities (e.g., clumpy
structure), as seen in the case of the Wolf-Rayet nebula NGC\,6888
\citep[][]{Stock2010,Toala2016a}. If clumps and filaments resultant
from instabilities are present in the Bubble Nebula, their effects in
mixing outer ionised material into the hot bubble must be small so
that the X-ray emissivity is still below the detection limit of the
current X-ray satellites.  We can estimate upper limits to the
electron densities of the hot gas in the Bubble Nebula by using the
normalisation parameters obtained with PIMMS (see Section~4.1).
Adopting a plasma temperature of 0.22~keV and 2.2~keV \citep[as those
  predicted by][]{Green2019}, we estimate upper limits of
$n_\mathrm{e}<$0.07~cm$^{-3}$ and $n_\mathrm{e}<$0.04~cm$^{-3}$, which
are very low compared to the detections of soft X-ray emission from
other WBBs.  \citet{Green2019} argue that some structures around
runaway stars, in particular hydrodynamical instabilities, might be
suppressed by magnetic fields.  This would certainly reduce the
efficiency of mixing into the hot bubble causing a low X-ray flux, but
we note that magnetic fields have not been reported to be present in
the Bubble Nebula so far.

The kinematic structure of the Bubble Nebula can provide an
alternative explanation to the lack of diffuse X-ray emission.
\citet{Chris1995} suggested a broken bubble-like structure based on
limited kinematic data, as could be expected in \citet{Green2019}'s
models, but the larger coverage of our SPM-MES observations unveils
that NGC\,7635 is actually composed by several disrupted bubble-like
structures or blisters.  In particular, a bright cavity with an
angular extension of $\sim40''$ ($\approx$0.50~pc) seems to surround
BD$+$60$^{\circ}$2522, whereas an even larger cavity extends mostly
towards the southern regions of the Bubble Nebula with an extension of
$\sim120''$ ($\approx$1.45~pc), as shown in the top panel of
Figure~10.  The Bubble Nebula is not ``a favourably oriented dense bow
shock'', but a series of nested shells that joint together into an
apparently single bubble, which are not expected in
\citet{Green2019}'s models, although projection effects result in
apparent closed shells in their synthetic H$\alpha$ and 24~$\mu$m
images.  The physical structure of the Bubble Nebula might be
indicative of its complex interaction with its surroundings, which may
involve different episodes of enhanced mass-loss from the central star
\citep[as suggested by the larger nested shells detected in wide-field
  optical images; e.g.,][]{Moore2002} or successive interactions of a
moving star with a stratified ISM in addition to the growth of
hydrodynamical instabilities \citep[][]{Pittard2013}.

The non-detection of extended X-ray emission filling the Bubble Nebula
is in line with the growing number of bow shocks not detected in X-rays. 
So far, 13 runaway hot stars have been studied with X-ray observations, 
but only one, namely $\zeta$~Oph, has resulted in a positive detection 
of extended thermal X-ray emission
\citep{Toala2016b,Toala2017b,DeBecker2017}. $\zeta$~Oph exhibits a
clear bow shock in IR and optical, but lacks a complete (close)
bubble-like morphology. Nevertheless, the extended (thermal) X-rays
are detected at the wake of the bow shock as predicted by numerical
simulations \citep[see ][]{Mackey2015,Green2019}.

\section{Summary} 
\label{sec:summary}

We have presented a multiwavelength study of the Bubble Nebula
(a.k.a. NGC\,7635) and its central star, BD$+$60$^{\circ}$2522. We
have used the stellar atmosphere code PoWR to characterise in
unprecedented detail the Ofp-type star BD$+$60$^{\circ}$2522. Our
best-fit model as well as the results from its X-ray emission shows
that this star is a classic young O-type star. Its stellar wind
parameters are $v_{\infty}$=2000~km~s$^{-1}$ and
$\dot{M}$=1.3$\times$10$^{-6}$~M$_{\odot}$~yr$^{-1}$, a stellar mass
of 27$\pm$7~M$_{\odot}$ with abundances very close to those of the
Sun, in accordance to previous observational estimates. We found that
BD$+$60$^{\circ}$2522 is X-ray bright with a dominant plasma
temperature of 0.60~keV (=$7\times10^{6}$~K) and an X-ray luminosity
in the 0.4--3.0~keV energy range that fulfills the luminosity criteria
of $L_\mathrm{X}/L_\mathrm{bol}\approx$10$^{-7}$. Our stellar
atmosphere model of BD$+$60$^{\circ}$2522 has been improved by
including the X-ray properties of this star. In particular, the
observed P Cygni line profile of the N\,{\sc v} resonance doublet can
only be reproduce by Auger ionisation due to the X-ray emission from
the star.

Although it is clear that the wind from BD$+$60$^{\circ}$2522 is
powerful enough to produce the Bubble Nebula, no diffuse X-ray
emission is detected within this WBB. Assuming that the Bubble Nebula
is filled with hot gas at $kT$=0.22~keV we estimated an upper limit to
the X-ray flux and luminosity of
$F_\mathrm{soft,X}<$1.0$\times$10$^{-14}$~erg~cm$^{-2}$~s$^{-1}$ and
$L_\mathrm{soft,X}<9.0\times10^{30}$~erg~s$^{-1}$, respectively. These
estimates are even lower if hotter gas is adopted. The estimated upper
limit of electron density $n_\mathrm{e}<$0.04--0.07~cm$^{-3}$ suggests
that mixing between the hot bubble and the outer nebular material has
not been efficient enough -- or it has been supressed -- to produce
detectable emissivity values or that the Bubble Nebula has a different
origin.

Kinematic data unveil the presence of a series of of nested shells
that joint together into an apparently single bubble, with additional
blisters that have grown to the point that have disrupted some regions
of NGC\,7635.  This structure, in conjunction with the notable space
velocity of BD$+$60$^{\circ}$2522, support the idea that the nebula
formed as the result of the motion of the star and the interaction of
successive episodes of enhanced mass-loss with a stratified
medium. While the model proposed by \citet{Green2019} is interesting,
it is not clear that it includes all ingredients required for a direct
comparison with the observations. Further improvements, such as
accounting for density gradients, magnetic fields, and successive
episodes of enhanced mass-loss, might be necessary to fully understnd
the complexity of the Bubble Nebula.

The search for thermal X-ray observations of runaway stars and their
bow shocks has resulted in negative reports for most cases except for
$\zeta$~Oph. Thus, the lack of diffuse X-ray emission from the Bubble
Nebula seems to be the rule and not the exception.

\section*{Acknowledgements}
% Entry for the table of contents, for this guide only
\addcontentsline{toc}{section}{Acknowledgements}

The authors thank the referee for a useful report that improved the
presentation of this paper. This work is based on observations
obtained with {\it XMM-Newton}, an ESA science mission with
instruments and contributions directly funded by ESA Member States and
NASA. The NOT data presented were obtained with ALFOSC and FIES, which
are provided by the Instituto de Astrof\'{i}sica de Andaluc\'{i}a
(IAA-CSIC) under a joint agreement with the University of Copenhagen
and NOTSA. This paper makes use of observations carried out at the
Observatorio Astron\'{o}mico Nacional on the Sierra San Pedro
M\'{a}rtir (OAN-SPM), Baja California, Mexico. We thank the daytime
and night support staff at the OAN-SPM for facilitating and helping
obtain our observations. JAT, MAG and HT are funded by UNAM DGAPA
PAPIIT project IA100318. GRL acknowledges support from Fundaci\'{o}n
Marcos Moshinsky, CONACyT and PRODEP (Mexico).  MAG acknowledges
support of the grants AYA 2014-57280-P and PGC2018-102184-B-I00 of the
Spanish Ministerio de Ciencia, Innovaci\'on y Universidades, cofunded
with FEDER funds. LS is supported by UNAM-PAPIIT grant IN101819.

%%%%%%%%%%%%%%%%%%%%%%%%%%%%%%%%%%%%%%%%%%%%%%%%%%



%\bibliographystyle{mnras}

\begin{thebibliography}{99}

\bibitem[Anand et al.(2009)]{Anand2009} Anand, M.~Y., Kagali, B.~A.,
  \& Murthy, J.\ 2009, Bulletin of the Astronomical Society of India,
  37, 1
  
\bibitem[Arthur(2012)]{Arthur2012} Arthur, S.~J.\ 2012, \mnras, 421,
  128 3

\bibitem[Bailer-Jones et al.(2018)]{BJ2018} Bailer-Jones, C.~A.~L.,
  Rybizki, J., Fouesneau, M., et al.\ 2018, \aj, 156, 58
  
\bibitem[Cardelli et al.(1989)]{Cardelli1989} Cardelli, J.~A.,
  Clayton, G.~C., \& Mathis, J.~S.\ 1989, ApJ, 345, 245

\bibitem[Christopoulou et al.(1995)]{Chris1995} Christopoulou, P.~E.,
  Goudis, C.~D., Meaburn, J., Dyson, J.~E., \& Clayton, C.~A.\ 1995,
  \aap, 295, 509

\bibitem[De Becker et al.(2017)]{DeBecker2017} De Becker, M., del
  Valle, M.~V., Romero, G.~E., Peri, C.~S., \& Benaglia, P.\ 2017,
  \mnras, 471, 4452
  
\bibitem[Dwarkadas \& Rosenberg(2013)]{Dwarkadas2013} Dwarkadas,
  V.~V., \& Rosenberg, D.~L.\ 2013, High Energy Density Physics, 9,
  226
  
%\bibitem[Freeman et al.(2014)]{Freeman2014} Freeman, M., Montez, R.,
%  Jr., Kastner, J.~H., et al.\ 2014, \apj, 794, 99
  
\bibitem[Fang et al.(2016)]{Fang2016} Fang, X., Guerrero, M.~A.,
  Toal{\'a}, J.~A., Chu, Y.-H., \& Gruendl, R.~A.\ 2016, \apjl, 822,
  L19
  
\bibitem[Gaia Collaboration et al.(2016)]{Gaia2016} Gaia
  Collaboration, Brown, A.~G.~A., Vallenari, A., et al.\ 2016, \aap,
  595, A2

\bibitem[Gaia Collaboration et al.(2018)]{Gaia2018} Gaia
  Collaboration, Brown, A.~G.~A., Vallenari, A., et al.\ 2018, \aap,
  616, A1

\bibitem[Gosset et al.(2005)]{Gosset2005} Gosset, E., Naz{\'e}, Y.,
  Claeskens, J.-F., et al.\ 2005, \aap, 429, 685
  
\bibitem[Gr{\"a}fener et al.(2002)]{Grafener2002} Gr{\"a}fener, G.,
  Koesterke, L., \& Hamann, W.-R.\ 2002, A\&A, 387, 244

\bibitem[Green et al.(2019)]{Green2019} Green, S., Mackey, J.,
  Haworth, T.~J., Gvaramadze, V.~V., \& Duffy, P.\ 2019,
  arXiv:1903.05505

\bibitem[Groenewegen, \& Lamers(1989)]{Groenewegen1989} Groenewegen,
  M.~A.~T., \& Lamers, H.~J.~G.~L.~M.\ 1989, \aaps, 79, 359

%\bibitem[Gruendl et al.(2004)]{Gruendl2004} Gruendl, R.~A., Chu,
%  Y.-H., \& Guerrero, M.~A.\ 2004, \apjl, 617, L127
  
\bibitem[Gruendl et al.(2004)]{Gruendl2004} Gruendl R.~A., Chu Y.-H.,
  Guerrero M.~A., 2004, \apjl, 617, L127

\bibitem[G{\"u}del et al.(2008)]{Gudel2008} G{\"u}del, M., Briggs,
  K.~R., Montmerle, T., et al.\ 2008, Science, 319, 309

\bibitem[Hamann \& Gr{\"a}fener(2004)]{Hamann2004} Hamann, W.-R., \&
  Gr{\"a}fener, G.\ 2004, A\&A, 427, 697

\bibitem[Hamann et al.(2006)]{Hamann2006} Hamann, W.-R., Gr{\"a}fener,
  G., \& Liermann, A.\ 2006, \aap, 457, 1015
  
\bibitem[Howarth et al.(1997)]{Howarth1997} Howarth, I.~D., Siebert,
  K.~W., Hussain, G.~A.~J., \& Prinja, R.~K.\ 1997, \mnras, 284, 265

\bibitem[Hubrig et al.(2015)]{hubrig2015} Hubrig, S., Sch{\"o}ller,
  M., Kholtygin, A.~F., et al.\ 2015, \mnras, 447, 1885
  
\bibitem[Jaskot et al.(2011)]{Jaskot2011} Jaskot, A.~E., Strickland,
  D.~K., Oey, M.~S., Chu, Y.-H., \& Garc{\'{\i}}a-Segura, G.\ 2011,
  \apj, 729, 28

\bibitem[Mackey et al.(2015)]{Mackey2015} Mackey, J., Gvaramadze,
  V.~V., Mohamed, S., \& Langer, N.\ 2015, A\&A, 573, A10

\bibitem[Maucherat \& Vuillemin(1973)]{Maucherat1973} Maucherat,
  A., \& Vuillemin, A.\ 1973, \aap, 23, 147
  
\bibitem[Megier et al.(2009)]{Megier2009} Megier, A., Strobel, A.,
  Galazutdinov, G.~A., \& Kre{\l}owski, J.\ 2009, \aap, 507, 833
  
\bibitem[Mernier \& Rauw(2013)]{Mernier2013} Mernier, F., \& Rauw, G.\
  2013, NewA, 20, 42

\bibitem[Moore et al.(2002)]{Moore2002} Moore, B.~D., Walter, D.~K.,
  Hester, J.~J., et al.\ 2002, AJ, 124, 3313

\bibitem[Nebot G{\'o}mez-Mor{\'a}n \&
  Oskinova(2018)]{Nebot2018} Nebot G{\'o}mez-Mor{\'a}n, A.,
  \& Oskinova, L.~M.\ 2018, \aap, 620, A89
  
\bibitem[Pallavicini et al.(1981)]{Pallavicini1981} Pallavicini, R.,
  Golub, L., Rosner, R., et al.\ 1981, \apj, 248, 279

\bibitem[Pittard(2013)]{Pittard2013} Pittard, J.~M.\ 2013, \mnras,
  435, 3600

\bibitem[Prinja et al.(1990)]{Prinja1990} Prinja, R.~K., Barlow,
  M.~J., \& Howarth, I.~D.\ 1990, \apj, 361, 607

\bibitem[Ram{\'\i}rez-Ballinas et al.(2019)]{RamirezBallinas2019}
  Ram{\'\i}rez-Ballinas, I., Reyes-Iturbide, J., Toal{\'a}, J.~A., et
  al.\ 2019, \apj, 885, 116
  
\bibitem[Rauw et al.(2003)]{Rauw2003} Rauw, G., De Becker, M., \&
  Vreux, J.-M.\ 2003, \aap, 399, 287
  
\bibitem[Ruiz et al.(2013)]{Ruiz2013} Ruiz, N., Chu, Y.-H., Gruendl,
  R.~A., et al.\ 2013, \apj, 767, 35

\bibitem[Sander et al.(2015)]{sander2015} Sander, A.,
  Shenar, T., Hainich, R., et al.\ 2015, \aap, 577, A13

\bibitem[Seaton(1979)]{Seaton1979} Seaton, M.~J.\ 1979, \mnras,
  187, 73

\bibitem[Shenar et al.(2015)]{shenar2015} Shenar, T., Oskinova,
  L., Hamann, W.-R., et al.\ 2015, \apj, 809, 135

\bibitem[Shenar et al.(2014)]{shenar2014} Shenar, T., Hamann, W.-R.,
  \& Todt, H.\ 2014, \aap, 562, A118

\bibitem[Stock \& Barlow(2010)]{Stock2010} Stock, D.~J., \& Barlow,
  M.~J.\ 2010, \mnras, 409, 1429
  
\bibitem[Telting et al.(2014)]{Telting2014} Telting, J.~H., Avila, G.,
  Buchhave, L., et al.\ 2014, Astronomische Nachrichten, 335, 41

\bibitem[Toal{\'a} \& Arthur(2011)]{Toala2011} Toal{\'a}, J.~A., \&
  Arthur, S.~J.\ 2011, \apj, 737, 100

\bibitem[Toal{\'a} \& Arthur(2018)]{Toala2018} Toal{\'a}, J.~A., \&
  Arthur, S.~J.\ 2018, \mnras, 478, 1218

\bibitem[Toal{\'a} et al.(2018)]{Toala2018b} Toal{\'a}, J.~A.,
  Oskinova, L.~M., Hamann, W.-R., et al.\ 2018, \apjl, 869, L11
  
\bibitem[Toal{\'a} et al.(2017)]{Toala2017} Toal{\'a}, J.~A., Marston,
  A.~P., Guerrero, M.~A., Chu, Y.-H., \& Gruendl, R.~A.\ 2017a, \apj,
  846, 76

\bibitem[Toal{\'a} et al.(2017)]{Toala2017b} Toal{\'a}, J.~A.,
  Oskinova, L.~M., \& Ignace, R.\ 2017b, \apjl, 838, L19
  
\bibitem[Toal{\'a} et al.(2016a)]{Toala2016a} Toal{\'a}, J.~A.,
  Guerrero, M.~A., Chu, Y.-H., et al.\ 2016a, \mnras, 456, 4305

\bibitem[Toal{\'a} et al.(2016b)]{Toala2016b} Toal{\'a}, J.~A.,
  Oskinova, L.~M., Gonz{\'a}lez-Gal{\'a}n, A., et al.\ 2016b, ApJ,
  821, 79
  
\bibitem[Toal{\'a} et al.(2015)]{Toala2015} Toal{\'a}, J.~A.,
  Guerrero, M.~A., Chu, Y.-H., \& Gruendl, R.~A.\ 2015, MNRAS, 446,
  1083

\bibitem[Toal{\'a} \& Guerrero(2013)]{Toala2013} Toal{\'a}, J.~A., \&
  Guerrero, M.~A.\ 2013, \aap, 559, A52
  
\bibitem[Toal{\'a} et al.(2012)]{Toala2012} Toal{\'a},
  J.~A., Guerrero, M.~A., Chu, Y.-H., et al.\ 2012, \apj, 755, 77

\bibitem[Toal{\'a} \& Arthur(2011)]{Toala2011} Toal{\'a},
  J.~A., \& Arthur, S.~J.\ 2011, \apj, 737, 100

\bibitem[Tody(1993)]{Tody1993} Tody, D.\ 1993, Astronomical Data
  Analysis Software and Systems II, 173
  
\bibitem[Townsley et al.(2014)]{Townsley2014} Townsley, L.~K., Broos,
  P.~S., Garmire, G.~P., et al.\ 2014, \apjs, 213, 1

\bibitem[Weaver  et  al.(1977)]{Weaver1977}  Weaver, R.,  McCray,  R.,
  Castor, J., Shapiro, P., \& Moore, R.\ 1977, \apj, 218, 377


\end{thebibliography}


%%%%%%%%%%%%%%%

\end{document}